\documentclass[10pt, a4]{article}
\usepackage{hyperref}
\usepackage{fontenc}
\usepackage{lmodern}
\usepackage{microtype}
\usepackage{amsmath, amssymb, amsfonts}
\usepackage{booktabs}
\usepackage{graphicx}
\usepackage{multirow}
\usepackage{float}
\usepackage{authblk}
\usepackage{geometry}[margins=4cm]
\usepackage{lipsum}
\title{\textbf{Modeling smooth and localized mortality patterns across age, time, and space to uncover small-area inequalities}}
\date{}
\author[1,2]{Jacob Martin\thanks{Email: \href{mailto:jmartin267@ikasle.ehu.eus}{jmartin267@ikasle.ehu.eus}}}
\author[2]{Carlo Giovanni Camarda}
\affil[1]{Universidad del País Vasco (UPV/EHU)}
\affil[2]{Institut national d'études démographiques}
\usepackage[backend=biber, style=authoryear]{biblatex}
\begin{document}
	
	\maketitle
	\begin{abstract}
		Small-area mortality estimation is inherently difficult, as random fluctuations from low death counts can obscure real geographic differences. We introduce a flexible model that borrows strength across age, space, and time to estimate mortality schedules and trends in very small populations. The approach ensures smooth patterns across these dimensions while allowing localized breaks from the spatial structure, capturing broad trajectories as well as sharp local contrasts. We implement our model within a Penalized Spline framework and estimate it using Generalized Linear Array Model techniques, resulting in a computationally fast, interpretable, and parsimonious method. Crucially, it can readily incorporate sudden mortality shocks, such as the Covid-19 pandemic, making it highly versatile for real-world demographic and epidemiological challenges. We demonstrate its application by estimating life expectancy and age-specific mortality inequalities in over 4,800 small areas across the Greater London Authority from 2002 to 2024. 
	\end{abstract}

	\section{Challenges in modeling mortality in small areas}
    Space powerfully shapes human mortality. Even where overall mortality is low, striking inequalities persist across territories, communities, and neighborhoods. These spatial arise from and amplify socioeconomic divides, producing profound disparities in health outcomes. Studying mortality at fine spatial scales is essential: it uncovers localized hazards hidden in broader averages and reveals the true depth of inequality. For example, a single metropolitan area may present very large differences in mortality conditions between and even within neighborhoods.

    A fundamental problem undermines the study of small-area differences in age-specific mortality: small populations leading to large fluctuations in observed death rates. As the spatial resolution increases, the population at risk of death in each area decreases, and the observed age- and period-specific death counts will produce a very noisy signal of the underlying risk of death. In very small populations, a large proportion of observed age-specific death counts will be zero, not necessarily reflecting a very low mortality risk. To overcome this difficulty, one solution is to just group ages, time periods, and areas, but this decreases the granularity of estimates. For example, the United Kingdom Office for National Statistics (ONS) publishes estimates of period life expectancy from mortality rates grouped by 5-year age group, 3-year period, and lower tier local authority in England \parencite{ONS2024}. These estimates already show substantial spatial variation in life expectancy, but they cannot identify sub-municipality or sub-neighborhood inequalities and are also limited by their temporal resolution.
    
    Trying to overcome the fundamental problem of small populations or samples has a long tradition in statistics. In the context of survey statistics, small area estimation techniques are employed when estimates are needed for a smaller sub-sample the survey (be it a geographic area or a demographic category) than the survey was designed for \parencite{Ghosh2020}. Synthetic estimators reduce the variance by borrowing information from a larger area containing the smaller area of interest. Model-based estimators \parencite{FayHerriot1979} use covariates and area-specific random effects to account for both sampling variation and unexplained differences between areas. While our problem differs, as we are not working with survey data, we can use similar ideas in modeling mortality, as we will see below.
    
    Spatial epidemiologists have also developed modeling techniques to address the small area problem. For example, estimating disease incidence poses an essentially identical problem to that of mortality: even with complete registration of new cases of a disease, observed incidence rates in small areas will exhibit large stochastic fluctuations, which obscure the underlying spatial heterogeneity in disease risk. Especially in a Bayesian context, models for disease mapping use spatial \parencite{BesagYorkMollie1991, BernarndinelliMontomoli1992} and temporal \parencite{Bernarndinelli1995} strength-borrowing techniques. These approaches are based on the principle that neighboring areas and adjacent time points tend to have a more similar level of disease risk compared to those that are farther apart. However, spatial oversmoothing can be a risk when using space to borrow strength; the model of \textcite{BesagYorkMollie1991} included unstructured area-specific effects to counteract this. \textcite{Congdon2009} and \textcite{Congdon2014} have leveraged ideas and methods from this literature to model all-cause death counts. \textcite{BennettEtAl2023} and \textcite{RashidEtAl2021} leveraged a Bayesian hierarchical modeling framework for modeling small-area mortality trends in England \& Wales and London specifically. 

    More recently, demographers have developed methods that leverage regularities in the age pattern of human mortality to address these challenges, sometimes incorporating spatio-temporal modeling \parencite{Denecke2023}. Essentially, these approaches involve estimating or constructing a standard mortality schedule and modeling deviations from it. Some methods treat mortality in a given area as additive deviation from the standard \parencite{GonzagaSchmertmann2016}, 
    while others penalize deviations that differ significantly from the standard pattern \parencite{Schmertmann2021}. Another approach expresses the mortality schedule in a given area as a linear combination of the principal components of the standard \parencite{AlexanderZagheniBarbieri2017}. More generally the idea of invoking a standard schedule to model mortality in data-poor settings has a long tradition in demography. A well-known example is the use of the Brass relational logit model and model life tables \parencite{BrassScaleMortRelMod1971, ZabaRelMod1979}.

    Both spatial epidemiological and demographic approaches have notable drawbacks. Methods from spatial epidemiology often do not fully account for the age dimension, which is fundamental in human mortality, nor its interaction with space and time. On the other hand, while demographic approaches place age at the center of their approaches, they introduce challenges by relying on an external age standard. This standard is typically estimated from observed data, such as the average age-specific mortality rates for multiple countries or years, but the uncertainty in its estimation is never incorporated into the model. Additionally, both epidemiological and demographic methods are often implemented in a hierarchical Bayesian framework. While this approach offers advantages, it also requires computationally intensive Markov Chain Monte Carlo (MCMC) algorithms for estimation.
	
    In this article we propose a flexible data-driven model for mortality estimation in small areas that borrows strength across age, space, and time. The model consists of three interpretable components: (1) the overall age-time pattern of mortality across areas, (2) smooth deviations in age, space, and time, and (3) unsmooth terms that allow for localized breaks from the underlying spatial pattern. We estimate our model in Penalized Generalized Linear Model framework using fast Generalized Linear Array Model \parencite[GLAM]{Currie2006} techniques. Smooth functions over age, time, and space are modeled using penalized splines ($P$-splines) resulting in a natural extension of age-time $P$-spline smoothing that is widely used in mortality research \parencite{CurrieDurbanEilersSmoothForecast2004, Camarda2008, Camarda2019}. We illustrate our model by analyzing spatial variations in mortality across more than 4,800 very small areas within the Greater London Authority from 2002 to 2024. Previous research has examined mortality in small areas of England \& Wales, notably \textcite{RashidEtAl2021} which studied all of England \& Wales and \textcite{BennettEtAl2023} which studied London specifically. However, both studies stopped their analysis at 2019 and employed models that lacked flexibility in the temporal dimension, imposing linear trends. While \textcite{BennettEtAl2023} studied trends in life expectancy in London at the same spatial scale we do, they did not examine trends in age-specific differences. Furthermore, to our knowledge, neighborhood inequalities in COVID-19 mortality in London remain unstudied, and post-COVID-19 mortality trends have yet to be analyzed. Thus, this paper pursues two overarching aims: one methodological and one substantive. Methodologically, we develop a novel statistical model that enables flexible estimation of mortality in small areas. Substantively, our demographic and epidemiological objectives are to:  
    \begin{enumerate}
        \item Estimate small-area trends in age-specific mortality risks in London.  
        \item Examine inequalities in age-specific mortality and life expectancy over the 2002--2024 period.  
        \item Assess the impact of COVID-19 on intra-urban mortality inequalities in London.  
        \item Characterize post-COVID-19 mortality trends in London up to 2024.  
    \end{enumerate}

    The paper is structured as follows. In Section~\ref{sec:data} we discuss the context of mortality trends in England \& Wales and London, and we present the small-area mortality data for the Greater London Authority. In Section~\ref{sec:method}, we outline our modeling strategy. We present results for life expectancy estimates and trends in age-specific mortality in Section~\ref{sec:results}, and we discuss them and our model in Section~\ref{sec:discussion}. Finally, we conclude in Section~\ref{sec:conclusion}.
    
    \section{Data and context}\label{sec:data}

    \subsection{Mortality inequalities and trends: Evidence from England \& Wales, and London}

    Spatial inequalities in mortality in England \& Wales have become increasingly relevant in light of recent trends in mortality in the United Kingdom and Europe. Since 2000, mortality improvements in low-mortality countries have slowed compared to the pre-2000 period \parencite{Dowd2024}. In the UK, this slowdown has been particularly pronounced, leading to a widening gap in life expectancy between the UK and other high-income countries since 2011 \parencite{Polizzi2024, Leon2019, Hiam2020}. At the sub-national level, life expectancy has even declined in some constituent countries, local authorities, and Middle Super Output Areas (MSOAs, the third smallest statistical geography used by the ONS) \parencite{Hiam2020, Alexiou2021, RashidEtAl2021}. Given this stagnation and in some cases reversal of mortality progress, assessing the scale of spatial inequalities and identifying areas where trends are deteriorating is more critical than ever.

    Historically, mortality patterns in London have long garnered interest. Indeed, the demographic study of mortality traces its origins to John Graunt's \textit{Natural and Political Observations Made Upon the Bills of Mortality} in \citeyear{Graunt1662}. The modern scientific study of mortality inequalities in London dates back to the mid-20th century. \textcite{Hewitt1956}, drawing on data from the early 1950s, reported that mortality was highest in a cluster of boroughs in east-central London. Differences were most pronounced at midlife among men, with elevated mortality strongly associated with lower socioeconomic status, overcrowded housing, and, most notably, air pollution. In the 1980s, \textcite{Congdon1988} found higher mortality in east London wards, with differences highest in midlife ages. More recently, in their analysis of spatial mortality trends in all of England \& Wales, \textcite{RashidEtAl2021} estimated that some of the largest increases in life expectancy between 2002 and 2019 occurred in the London borough of Camden. In their London-specific analysis of Lower Super Output Areas (LSOAs, a statistical geography one level smaller than the MSOAs), \textcite{BennettEtAl2023} estimated that life expectancy was highest in 2019 in central and north-central London and lowest in the outer east and southeast. They further reported that the most substantial gains between 2002 and 2019 occurred in central and central eastern London, contributing to a widening gap in life expectancy across different areas. While both \textcite{RashidEtAl2021} and \textcite{BennettEtAl2023} examined mortality trends in London at the small-area level, their studies have two key limitations. First, their modeling strategy was somewhat rigid in the temporal dimension, as area specific mortality was assumed to follow a linear trend in time, and the models did not fully account for interaction between age, space, and time. Second, both analyses stop in 2019, leaving the short term and possibly lasting effects of Covid-19 on spatial mortality trends in London unexamined.

    \subsection{Small-area death and population data for London}
    
    We obtained mid-year population estimates and death counts from the Office for National Statistics (ONS) by single year of age, sex, calendar year, and Lower Super Output Area (LSOA) for all areas under the governance of the Greater London Authority (GLA) for the years 2002--2024. Population estimates were taken from a publicly available ONS dataset for 2002--2022 \parencite{ONSPopulation}, while deaths for the period 2002--2016 were obtained from another publicly available dataset \parencite{ONSDeaths2016}. For the years 2017--2024, death data by LSOA and single year of age were not publicly available and were therefore obtained through a direct request to the ONS. Finally, since the 2023 and 2024 LSOA-specific mid-year population estimates have not been published at the time of writing, we calculated LSOA population projections for 2023 and 2024 using the method outlined in the United Kingdom Department of Health and Social Care Public Health Technical Guidance \parencite{LSOAProjections}.
    
    The GLA is the regional governing body encompassing the London urban area. It consists of 33 local government districts: 32 boroughs and the City of London. In 2024, the GLA had an estimated total mid-year population of 9,089,736. LSOAs are a statistical geography defined by the ONS, each typically comprising a population between 1,000 and 3,000 residents. LSOA boundaries changed in 2011 and 2021, and not all data were available with consistent boundaries. Under the 2011 boundaries, London had 4,835 LSOAs, of which 4,659 remained unchanged. Merges and splits led to 4,994 LSOAs under the 2021 boundaries. To ensure consistent geographic units across the study period, we adopted the most granular grouping of LSOAs that maintained constant boundaries for all years, resulting in a total of 4,813 LSOAs for the GLA region. Throughout the analysis, we used death counts based on the LSOA of residence rather than the place of occurrence, to more accurately capture local population health patterns.
    
    For the population estimates, the open age category is 90+, so we set this as the open age group for the deaths as well. For a small proportion of cells in our dataset (74,580 cells, 0.37\% of the total number of cells), there were positive deaths but zero exposures, which is not possible and most likely due to estimation error in our population data. For these cells, we set the exposures equal to the number of deaths, and we subtracted a proportional amount from the exposures for the other ages in the same year, sex, and LSOA. This ensures that the LSOA population estimates sum to the same totals as in the original population estimates. We incorporate spatial information into our model by using the centroids of the LSOAs, calculated using the \texttt{sf} package in \texttt{R} \parencite{sfPackageBook2023} and the official LSOA boundary shapefile for England \& Wales (modified to reflect our consistent LSOA grouping). Thus, the data inputs we consider are: $y_{ijk}$, the number of deaths among individuals aged $i$ resident in LSOA $j$ during year $k$; $e_{ijk}$, the mid-year population aged $i$ in LSOA $j$ for mid-year $k$; and $(\mathrm{x,y})_j$ the longitude and latitude coordinates of the centroids of LSOA $j$. 
    
\begin{figure}
        \centering
        \includegraphics[width=\textwidth]{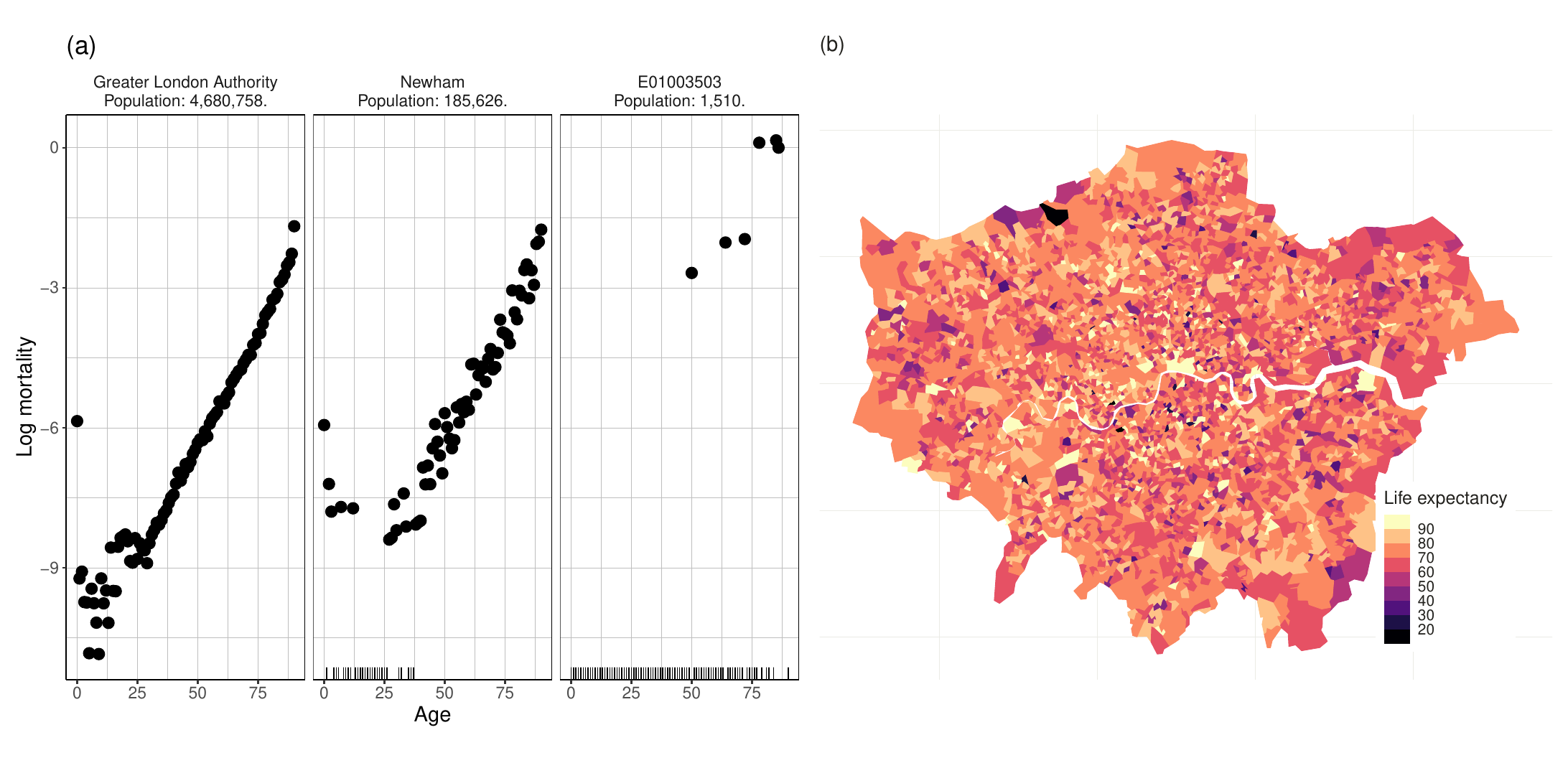}
        \caption{(a) Observed log-mortality rates for females in 2024 at three spatial aggregation levels. Tick marks along the horizontal axis represent ages with no deaths. \textbf{Left}: entire Greater London Authority (GLA). \textbf{Center}: London Borough of Newham. \textbf{Right}: single LSOA, Redbridge. The GLA shows a clearly distinguishable age pattern, with some fluctuations at younger ages due to fewer observed deaths. At the Borough level, many zeros occur at younger ages and the age pattern begins to break down, although mortality beyond age 50 still shows a strong signal. For the single LSOA, there are zero deaths at almost all ages, resulting in no identifiable mortality pattern.
        (b) Life expectancy at birth for females in 2024, calculated using observed mortality rates by LSOA. The estimated life expectancies span an implausibly wide range, from less than 7 years to over 90 years. Extremely high estimates occurred in areas where very few female deaths were recorded, resulting in an estimated force of mortality of zero across almost all ages. Conversely, the lowest life expectancy estimate arose in an LSOA where a single death at age 4 occurred among an estimated female population of only three, producing an anomalously high early-age mortality rate that substantially lowered the life expectancy calculation.}
        \label{fig:data}
    \end{figure}

    Panel (a) of Figure~\ref{fig:data} presents the observed age-specific mortality rates for females in 2024 at three levels of spatial aggregation, corresponding to different population sizes, plotted on a logarithmic scale. The observed log-mortality rates are calculated as 
    \[
    \eta^{\mathrm{obs}}_{ijk} = \ln\left(\frac{y_{ijk}}{e_{ijk}}\right)\,.
    \]
     Using the logarithmic scale is particularly convenient because human mortality risk varies over several orders of magnitude across different ages, allowing for a clearer visualization and interpretation of these wide-ranging values. Moreover, this transformation will prove natural in subsequent modeling stages, where log-mortality will be directly modeled.
     
     The clarity of the mortality signal diminishes as the population size decreases. At the highest level of aggregation, the entire GLA, with a female population exceeding four and a half million, the characteristic age pattern of mortality is distinct and readily identifiable. At the intermediate level, represented by the borough of Newham, which has a population of over 150,000, an increased number of zero death counts appear at younger ages, as indicated by the vertical tick marks along the horizontal axis, while greater variability is observed at older ages. Despite this, the overall pattern of adult mortality remains clearly discernible. In contrast, at the most granular level, a single Newham LSOA with approximately 1,000 residents exhibits extremely sparse deaths, resulting in a loss of any visually clear mortality signal.

    Given the difficulty of discerning meaningful age-specific mortality patterns and therefore of identifying real differences among small areas, life expectancy at birth may appear to offer a more stable and interpretable summary of mortality. Classically, life expectancy at birth represents the number of years lived by a hypothetical cohort exposed to all of the period age-specific mortality rates. It is also an extremely useful indicator that summarizes the whole age profile of mortality, since it does not depend on population age structure. We emphasize, however, that this measure should not be interpreted as the real average length of life in an area. Furthermore, at fine spatial scales, life expectancy can be highly unstable when calculated from directly observed mortality rates \parencite{Preston2001}, as shown in panel (b) of Figure~\ref{fig:data}.
    
    Areas with zero observed deaths across multiple ages will have unrealistically high life expectancy estimates. On the other hand, in areas with just one or two deaths at younger ages, where population exposures are small, life expectancy can be drastically underestimated, as a single death can translate into an inflated observed mortality rate. As shown in Figure~\ref{fig:data} panel (b), these issues result in life expectancy values ranging from below 7 years to over 90 years. While the extremely low values are clearly implausible, the overall magnitude of the range also fails to reflect any genuine heterogeneity between LSOAs. Indeed, the complete lack of any spatial pattern suggests the observed differences are driven entirely by random noise in the small death counts.

\section{A spatial $P$-spline model of mortality}\label{sec:method}

    \subsection{Model formulation}

        Our goal is to produce robust estimates of age-specific mortality risk at the LSOA level, disaggregated by single years of age and time period. To this end, we develop a model that accounts for interactions between age, time, and space, while borrowing strength across all areas under study. At the same time, we aim to avoid spatial oversmoothing: although borrowing information from neighboring areas helps reveal spatial patterns, we recognize that localized hazards can generate abrupt, non-smooth breaks in space. 
        
        We model males and females separately. For a given sex, let deaths and mid-year populations be denoted as follows. We have $m = 91$ ages, $n = 4813$ spatial units, and $l = 23$ years, so for each sex we have to model a total of 10,073,609 mortality rates. We arrange the deaths into a three-way array $\mathbf{Y}$ of dimension $m \times n \times l$. Similarly, we arrange the mid-year population exposures into a $m \times n \times l$ array $\mathbf{E}$. We assume that $y_{ijk} \sim \mathrm{Poisson}(\mu_{ijk} e_{ijk})$, where $\mu_{ijk}$ denotes the force of mortality at age $i$ in area $j$ during year $k$. Within the standard regression framework, the log-linear model for the expected number of deaths is given by
        \begin{equation}
        	\ln E[\mathbf{y}] = \ln(\mathbf{e}) + \ln(\boldsymbol{\mu}) = \ln(\mathbf{e}) + \boldsymbol{\eta} = \ln(\mathbf{e}) + \mathbf{X}\boldsymbol{\theta}\, ,
        \end{equation}
        where $\mathbf{y} = \text{vec}(\mathbf{Y})$ denotes the vectorized death counts, $\mathbf{e} = \text{vec}(\mathbf{E})$ the corresponding exposures (included as an offset), $\mathbf{X}$ the design matrix encoding covariates and basis functions (e.g., age, space, time), and $\boldsymbol{\theta}$ the vector of coefficients to be estimated. Specifically, we model log-mortality $\boldsymbol{\eta}_j$ for region $j$ as the sum of smooth and unsmooth components:
        \begin{equation}
        	\boldsymbol{\eta}_j = \boldsymbol{\eta}^0 + \boldsymbol{\delta}_j + \gamma_j,
        \end{equation}
        where $\boldsymbol{\eta}^0$ denotes a baseline schedule common to all regions, estimated smoothly over age and time; $\boldsymbol{\delta}_j$ represents age-specific deviations from this baseline, modeled to vary smoothly across age, space, and time, and $\gamma_j$ are unsmooth area-specific level terms. 
    
    \subsection{Representation of components in the model matrix}
    
        This model structure is reflected in the construction of the model matrix $\mathbf{X}$, which incorporates all relevant components: smooth terms over age and time common to all areas, as well as interactions between smooth terms over age, space, and time. All smooth components are modeled using penalized splines: we select a relatively large number of equally spaced cubic $B$-splines, with coefficients subsequently penalized using a discrete penalty.
        
        The common smooth age-time pattern $\boldsymbol{\eta}^0$ is modeled as the tensor product $\mathbf{B}_t \otimes \mathbf{B}_a$, where $\mathbf{B}_a$ and $\mathbf{B}_t$ are marginal $B$-spline bases over age and time, respectively, and $\otimes$ is the Kronecker product. Specifically, $\mathbf{B}_a$ is an $m \times c_\mathrm{a}$ basis of $c_\mathrm{a} =21$  $B$-splines. To account for the sharp drop in mortality after infancy, the coefficient for age 0 is modeled independently of the other ages \parencite{Camarda2019}. The time basis $\mathbf{B}_t$ is an $l \times c_\mathrm{t}$ $B$-spline basis with $c_\mathrm{t} = 7$. A linear combination of the tensor-product basis functions defines the smooth age-time pattern of mortality, a widely used approach for modeling mortality dynamics over age and time while ignoring spatial variation \parencite{CurrieDurbanEilersSmoothForecast2004, Camarda2008}. Since the age-time pattern is common for all areas, we expand it as $\mathbf{B}_t\otimes \mathbf{1}_{n}\otimes\mathbf{B}_a$, where $\mathbf{1}_n$ is a vector of ones of length $n$.
        
\begin{figure}
        	\centering
        	\includegraphics[width=\textwidth]{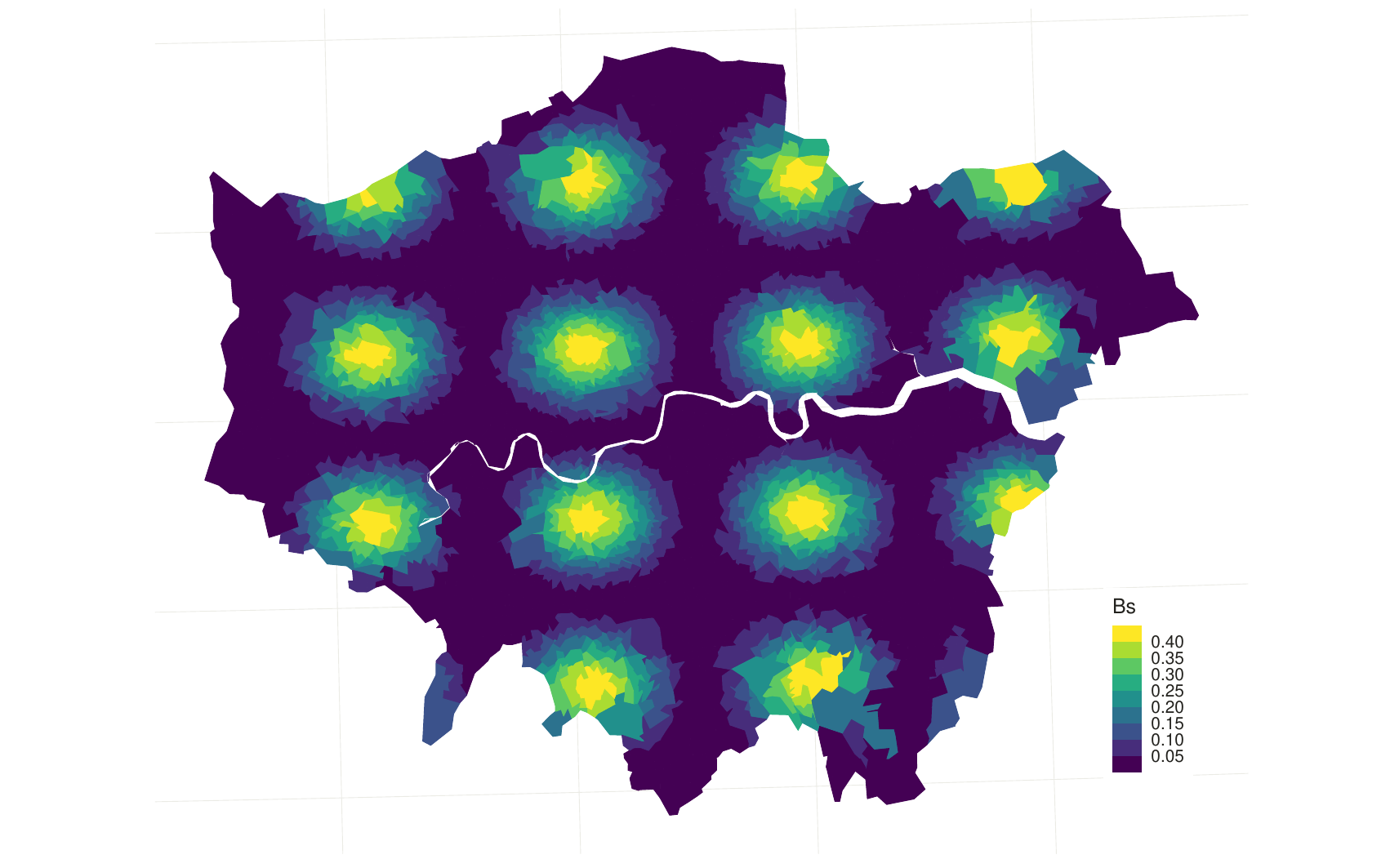}
        	\caption{$B$-spline basis functions over space. Only a subset of 121 bases is shown to reduce visual clutter and clearly convey the concept of the spatial basis.}
        	\label{fig:Bs}
        \end{figure}

        The deviations $\boldsymbol{\delta}_j$ are modeled to vary smoothly over age, space, and time. To capture the smooth spatial pattern, let $\mathbf{B}_\mathrm{lon}$ ($n \times c_\mathrm{lon}$) and $\mathbf{B}_\mathrm{lat}$ ($n \times c_\mathrm{lat}$) be the marginal $B$-spline bases over the longitude and latitude coordinates of the centroids, respectively, with 11 $B$-splines in each dimension. A smooth spatial basis $\mathbf{B}_s$ is then constructed via the row tensor, or ``box'' product of $\mathbf{B}_\mathrm{lon}$ and $\mathbf{B}_\mathrm{lat}$:
        \begin{equation}\label{eq:Bs}
        	\mathbf{B}_s = \mathbf{B}_\mathrm{lat} \Box \mathbf{B}_\mathrm{lon}  = (\mathbf{B}_\mathrm{lat} \otimes \mathbf{1}_{c_\mathrm{lon}}') \odot (\mathbf{1}_{c_\mathrm{lat}}' \otimes \mathbf{B}_\mathrm{lon}) \, ,
        \end{equation}
        where $\mathbf{1}_{c_\mathrm{lon}}$ and $\mathbf{1}_{c_\mathrm{lat}}$ are row vectors of ones of length $c_\mathrm{lon}$ and $c_\mathrm{lat}$, respectively; $\odot$ denotes element-wise multiplication, and the superscript $'$ indicates the transpose. The resulting $\mathbf{B}_s$ provides a smooth basis over both spatial dimensions \parencite{LeeDurban2011}, with dimension $n \times c_\mathrm{s}$, where $c_\mathrm{s} = c_\mathrm{lon} c_\mathrm{lat} = 121$. We use the box product rather than the Kronecker product because the centroid coordinates are scattered rather than on a regular grid \parencite{Eilers2006}. Figure~\ref{fig:Bs} illustrates a subset of the spatial bases over a map of London. To allow for interactions with age and time, we introduce reduced marginal bases $\mathbf{\breve{B}}_a$ ($m \times \breve{c}_\mathrm{a}$, adjusted for infant mortality) and $\mathbf{\breve{B}}_t$ ($l \times \breve{c}_\mathrm{t}$) and with fewer $B$-splines than $\mathbf{B}_a$ and $\mathbf{B}_t$, respectively. We use nested bases for $\breve{\mathbf{B}}_a$ and $\breve{\mathbf{B}}_t$ \parencite{Lee2013}, so that $\breve{c}_a = 9$ and $\breve{c}_t = 5$. The full age-space-time interaction component of the model matrix associated with $\boldsymbol{\delta}_j$ is then given by $\mathbf{\breve{B}}_t \otimes \mathbf{B}_s \otimes \mathbf{\breve{B}}_a$.
        
        The last component of our model are the $\gamma_j$, area-specific terms that are not smooth in space. With the previous model components, we have assumed the presence of an overall spatial structure in mortality, whereby geographically proximate areas tend to exhibit more similar rates than distant ones. However, relying solely on this smooth component risks oversmoothing and thereby masking meaningful local deviations. The inclusion of the non-smooth $\gamma_j$ terms allows the model to capture such localized departures from the general spatial trend, preserving area-specific risks that might otherwise be attenuated. For instance, while a neighborhood or cluster of areas may generally exhibit low mortality, the presence of an environmental hazard confined to a single area could produce a sharp increase in mortality that deviates from the surrounding pattern. 
        
        We assume that the unstructured area-specific effects are constant across age groups and represent them in the model matrix $\mathbf{X}$ using the $n \times n$ identity matrix $\mathbf{I}_n$, expanded as $\mathbf{1}_l \otimes\mathbf{I}_n \otimes \mathbf{1}_m$, where $\mathbf{1}_l$ and $\mathbf{1}_m$ are column vectors of ones of length $l$ and $m$, respectively. These effects are penalized as specified below to ensure identifiability of the overall model. This formulation is related to the Penalized Regression with Individual Deviance Effects (PRIDE) model \parencite{Perperoglou2010}; however, unlike PRIDE, we introduce deviance effects only at the level of the $n$ areas, rather than for each observation (i.e., each age–area-year-specific death count). This restriction preserves smoothness in the fitted mortality schedules across ages, which would otherwise be disrupted by observation-level deviances. It also yields a more parsimonious specification, requiring only $n$ $\gamma_j$ parameters instead of $mnl$, under the assumption that age-specific variation is largely captured by the smooth terms. Finally, given the sparsity of our data, meaningful residual unstructured variation is unlikely to be identifiable at individual ages, further motivating this simplification.
        
        The model matrix would thus take the form
        \[\mathbf{X} = \mathbf{B}_t \otimes \mathbf{1}_n\otimes\mathbf{B}_a: \mathbf{\breve{B}}_t\otimes\mathbf{B}_s \otimes \mathbf{\breve{B}}_a: \mathbf{1}_l\otimes\mathbf{I}_n\otimes\mathbf{1}_m,\]
        but this specification introduces identifiability issues between the components $\mathbf{B}_t \otimes \mathbf{1}_n\otimes\mathbf{B}_a$ and $\mathbf{\breve{B}}_t\otimes\mathbf{B}_s \otimes \mathbf{\breve{B}}_a$. For example, part of the baseline mortality level could be subtracted from the overall smooth age pattern $\boldsymbol{\eta}^0$ and equivalently absorbed into all smooth age-space deviations $\boldsymbol{\delta}_j$, making the two components indistinguishable without further constraints. To address this identifiability issue, we augment $\mathbf{Y}$ and $\mathbf{E}$ with an additional row representing the total deaths and exposures across all areas (i.e., for the entire GLA). We then correspondingly adjust the model matrix as follows:
        \begin{equation}\label{eq:ModelMatrix}
        	\renewcommand{\arraystretch}{1.4}
        	\mathbf{X} = 
        	\left[
        	\begin{array}{@{}c|c@{}}
        		\multirow{2}{*}{  $\mathbf{B}_t \otimes \mathbf{1}_{n+1} \otimes \mathbf{B}_{a} $ } & \mathbf{0}_{m , \breve{c}_{\mathrm{a}c_\mathrm{s}} +n\breve{c}_{\mathrm{t}}} \\
        		\cline{2-2}
        		& \breve{\mathbf{B}}_{t}\otimes\mathbf{B}_{s} \otimes  \breve{\mathbf{B}}_{a} \; \Big{|}\; \mathbf{1}_l \otimes \mathbf{I}_{n} \otimes \mathbf{1}_{m}
        	\end{array} \right].
        \end{equation}
        where $\mathbf{0}_{m , \breve{c}_{\mathrm{a}c_\mathrm{s}} +n\breve{c}_{\mathrm{t}}}$ is a matrix of zeros of dimension $ m \times ( \breve{c}_{\mathrm{a}c_\mathrm{s}} +n\breve{c}_{\mathrm{t}})$. Our final model matrix has a total of 12,271 columns.
    
    \subsection{Smoothing and estimation}
    
        Smoothness is enforced via a second-order difference penalty on the coefficients of the spline bases over age, space, and their interaction, while identifiability is maintained through a ridge penalty on the $\gamma_j$. Let $\mathbf{D}_a$, $(c_\mathrm{a} - 2) \times c_\mathrm{a}$ be the second difference matrix acting on the coefficients associated with $\mathbf{B}_a$, and let $\mathbf{D}_t, (c_\mathrm{t} - 2) \times c_\mathrm{t}$ be the difference matrix for $\mathbf{B}_t$. The penalty for smoothness over age and time for $\boldsymbol{\eta}^0$ is thus 
        \[\mathbf{P}_1 = \lambda_a\mathbf{I}_{c_\mathrm{t}}\otimes\mathbf{D}_a'\mathbf{D}_a + \lambda_t\mathbf{I}_{c_\mathrm{a}}\otimes\mathbf{D}_t'\mathbf{D}_t\] 
        where $\lambda_a$ is the smoothing parameter for age, $\lambda_t$ is the smoothing parameter for time. For the smooth age-space-time deviations $\boldsymbol{\delta}_j$, let $\mathbf{D}_\mathrm{lon}$, $\mathbf{D}_\mathrm{lat}$, $\mathbf{\breve{D}}_a$, and $\mathbf{\breve{D}}_a$ be the second difference matrices acting on the coefficients associated with $\mathbf{B}_\mathrm{lon}$, $\mathbf{B}_\mathrm{lat}$, $\mathbf{\breve{B}}_a$, and  $\mathbf{\breve{B}}_t$, respectively. The penalty over both spatial dimensions and age is thus
        \begin{align*}
        	\mathbf{P}_2 &= \lambda_{\mathrm{lon}}\mathbf{I}_{\breve{c}_\mathrm{t}}\otimes\mathbf{I}_{c_{\mathrm{lat}}}\otimes \mathbf{D}_\mathrm{lon}'\mathbf{D}_\mathrm{lon} \otimes \mathbf{I}_{\breve{c}_{\mathrm{a}}} \\ 
        	&+ \lambda_{\mathrm{lat}}\mathbf{I}_{\breve{c}_\mathrm{t}}\otimes \mathbf{D}_\mathrm{lat}'\mathbf{D}_\mathrm{lat}\otimes \mathbf{I}_{c_{\mathrm{lon}}} \otimes \mathbf{I}_{\breve{c}_{\mathrm{a}}} \\
        	&+\breve{\lambda}_a\mathbf{I}_{\breve{c}_\mathrm{t}}\otimes \mathbf{I}_{k_{\mathrm{lat}}}\otimes\mathbf{I}_{k_{\mathrm{lon}}} \otimes \breve{\mathbf{D}}_a'\breve{\mathbf{D}}_a \\
        	&+\breve{\lambda}_t \breve{\mathbf{D}}_t'\breve{\mathbf{D}}_t\otimes \mathbf{I}_{c_{\mathrm{lat}}}\otimes\mathbf{I}_{c_{\mathrm{lon}}} \otimes \mathbf{I}_{\breve{c}_\mathrm{a}},
        \end{align*}
        where $\lambda_{\mathrm{lon}}$, $\lambda_{\mathrm{lat}}$, $\breve{\lambda}_a$, and $\breve{\lambda}_t$ are the smoothing parameters for longitude, latitude, the second age basis, and the second time basis, respectively. Note that we require separate combinations of smoothing parameters, $(\lambda_a, \lambda_t)$ for the standard age-time schedule and $(\breve{\lambda}_a, \breve{\lambda}_t)$ for the age-time deviations, since the degree of smoothness for the deviations is expected to differ from that of the standard schedule. Finally, we penalize the $\gamma_j$ using a ridge penalty:
        \[\mathbf{P}_3 = \kappa \, \mathbf{I}_n;\]
        where $\kappa$ controls the magnitude of the shrinkage. The complete penalty matrix has then a block-diagonal structure:
        \[
        \mathbf{P} = \mathrm{diag}(\mathbf{P}_1, \mathbf{P}_2, \mathbf{P}_3)\, .
        \]
        
        With the model matrix $\mathbf{X}$ and the penalty $\mathbf{P}$ specified, the regression coefficients $\boldsymbol{\theta}$ are estimated via a penalized version of the iteratively re-weighted least squares (IRWLS) algorithm \parencite{McCullaghNelder1989}:
        \begin{equation} \label{eq:irwls}
            (\mathbf{X}'\tilde{\mathbf{W}}\mathbf{X}+\mathbf{P})\hat{\boldsymbol{\theta}} = \mathbf{X}'\tilde{\mathbf{W}}\tilde{\mathbf{z}},
        \end{equation}
        where a tilde denotes working values at each step of the algorithm, and $\hat{\boldsymbol{\theta}}$ are the updated coefficients at each iteration. As in conventional Poisson GLMs, the diagonal weight matrix $\tilde{\mathbf{W}} = \mathrm{diag}(\tilde{\boldsymbol{\mu}} \odot \mathbf{e})$ and the working dependent variable $\tilde{\mathbf{z}} = \tilde{\boldsymbol{\eta}} + \tilde{\mathbf{W}}^{-1} (\mathbf{y} - \tilde{\boldsymbol{\mu}} \odot \mathbf{e})$ are defined, respectively. An estimate of the variance-covariance matrix of the estimated parameters $\boldsymbol{\hat{\theta}}$ is given by $\mathbf{V = (\mathbf{X'WX+P})^{-1}}$ \parencite{McCullaghNelder1989}. To evaluate uncertainty around functions of the parameters like life expectancy, we can use a parametric bootstrap procedure, simulating parameters using a multivariate normal distribution with mean $\boldsymbol{\hat{\theta}}$ and variance-covariance $\mathbf{V}$.
        
        Applying the IRWLS algorithm in \eqref{eq:irwls} directly is computationally infeasible, as even constructing the model matrix would create storage issues. To address this, we leverage Generalized Linear Array Model (GLAM) techniques \parencite{Currie2006} wherever possible. This approach allows us to avoid explicitly computing the full Kronecker products in \eqref{eq:ModelMatrix}, while still enabling efficient computation of the inner products and linear functions in \eqref{eq:irwls} in both time and memory. We do, however, still need to compute the box product in \eqref{eq:Bs}. Although this represents a limitation, it is not prohibitive. Given the large number of spatial units, only a moderate number of basis functions is required in each spatial dimension \parencite{Ruppert2002}, and sufficient flexibility is additionally provided by the $\gamma_j$, ensuring that the box product in $\mathbf{B}_s$ remains computationally feasible.
        
        The smoothing parameters can $\lambda_a, \lambda_t, \breve{\lambda}_a, \breve{\lambda}_t, \lambda_\mathrm{lon},\lambda_\mathrm{lat}$ and the ridge penalty $\kappa$ can be selected through a grid search to optimize an information criterion. In our application we will optimize the Hannan–Quinn information criterion (HQIC) \parencite{HannanQuinn1979}, given by 
        \[
        \text{HQIC} = \text{DEV} + 2\cdot\ln(\ln(N))\cdot\text{ED},
        \]
        where $\text{DEV}$ is the deviance of the model fitted with a given choice of smoothing parameters in a Poisson setting, $\text{ED}$ is the effective dimension computed as the trace of the hat matrix at convergence of \eqref{eq:irwls}, and $N$ is the sample size.
        
        The HQIC provides a valid alternative to the more widely used Akaike Information Criterion (AIC) and Bayesian Information Criterion (BIC), particularly in settings with very sparse data, as in our application. The AIC is known to overfit mortality data, favoring overly flexible models \parencite{CurrieDurbanEilersSmoothForecast2004}, whereas the BIC, due to the large number of zero counts, tends to oversmooth by substantially reducing the deviance component in the criterion. The HQIC, which is strongly consistent like the BIC, uses the $2 \ln(\ln(N))$ penalty term to strike a balance, producing a degree of smoothness that lies between that selected by the AIC and BIC and is more appropriate for sparse mortality data.
        
        Given that we have six smoothing parameters and one ridge penalty, a full grid search would be computationally intensive. To reduce complexity, we first perform the grid search for the $(\lambda_a, \lambda_t)$ combination, corresponding to smoothness over age and time for the aggregate across all areas. Furthermore, since the basis $\mathbf{\breve{B}}_t$ has only five associated coefficients, we set the corresponding penalty $\breve{\lambda}_t$ to zero, as additional penalization would have minimal effect. Consequently, the effective grid search is reduced to four parameters, ensuring computational feasibility while preserving adequate model flexibility.
        
        Our analysis includes the Covid-19 pandemic, a shock expected to induce departures from the smooth age–space–time mortality pattern. To account for this, we extend the matrices $\mathbf{B}_t$ and $\mathbf{\breve{B}}_t$ by adding two columns corresponding to the years 2020 and 2021, effectively introducing parameters that model Covid-19 as an additive deviation from the underlying age-space-time trends. Although the longer-term effects of the pandemic on mortality remain uncertain, we consider only 2020 and 2021 as shock years. This choice is supported by data indicating that weekly all-cause deaths in London returned close to the 2015--2019 average by 2022 \parencite{GLACovid}. A key advantage of including additive terms for 2020 and 2021 is that the long-term smooth trend is preserved across the entire 2002--2024 period. Importantly, the model retains sufficient flexibility to accommodate the possibility of elevated mortality beyond 2021; should the pandemic exert longer-lasting effects, the smooth component can capture deviations in 2022--2024 relative to 2019 levels.

        Finally, we provide the codes (but not the data as for some years access is restricted) that we used to produce the analysis in the following anonymized OSF repository: \\ \href{https://osf.io/bwe6s/overview?view_only=6b29d0f4d7d14eeb95b6828d5cc07fd4}{[https://osf.io/bwe6s/overview?view\_only=6b29d0f4d7d14eeb95b6828d5cc07fd4]}
    			
	\section{Results}\label{sec:results}
	
        \subsection{Life expectancy trends}
	
    	For context, directly estimated life expectancy for the whole of London increased from 80.9 years in 2002 to 85.1 years in 2024 for females, and from 76.1 years to 80.8 years for males. These increases were not constant: in correspondence of the Covid-19 pandemic life expectancy declined sharply from 84.7 in 2019 to 83.2 in 2020 for females and from 80.7 to 78.4 for males. Thus, life expectancy did not fully recover to pre-Covid-19 levels until 2024.  
    	
\begin{figure}
    		\centering
    		\includegraphics[width=\textwidth]{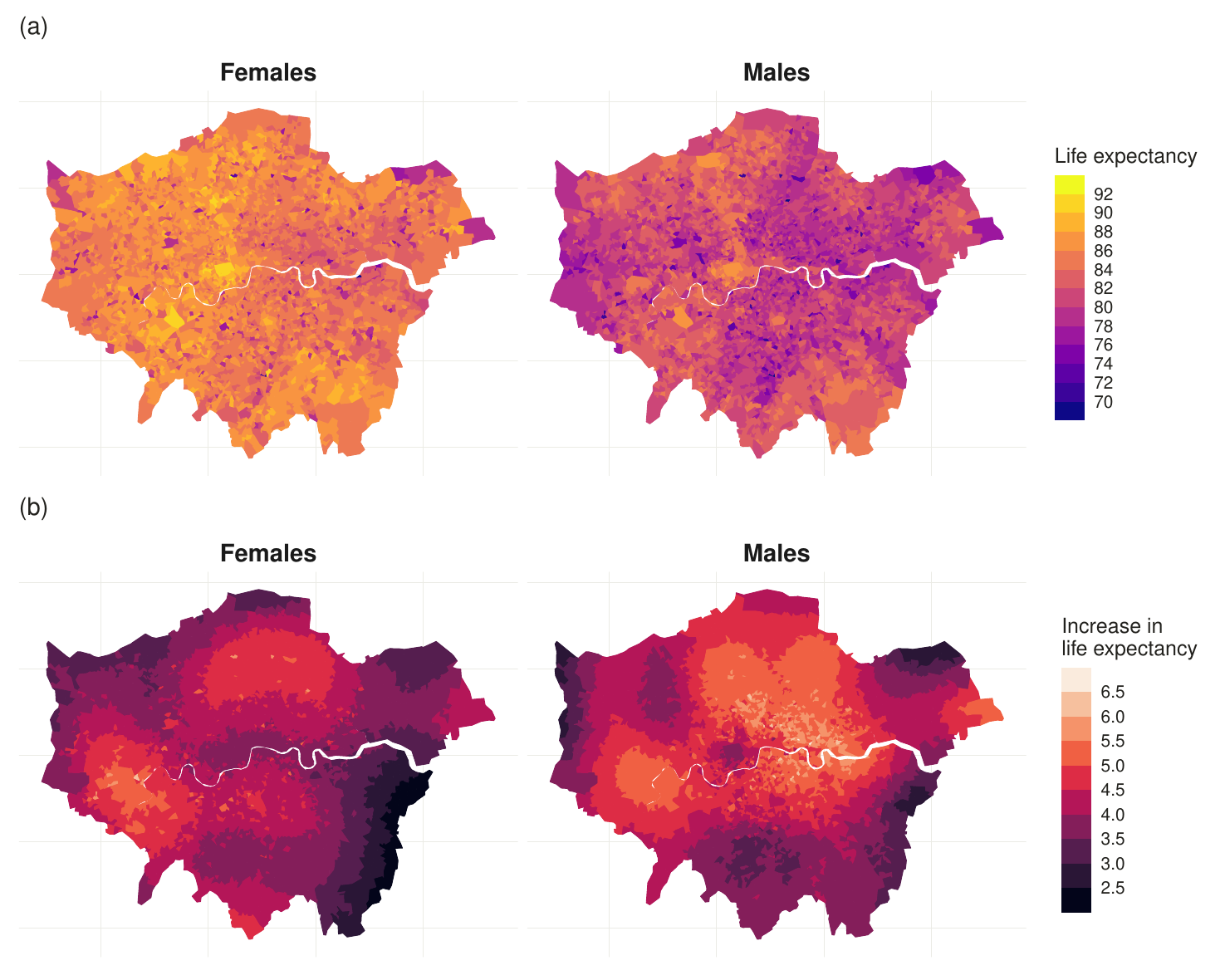}
    		\caption{(a) Estimated life expectancy at birth for all London LSOAs in 2024. For both males and females, the life expectancies were mostly clustered in the boroughs of Barnet and Kensington and Chelsea. The lowest life expectancy LSOAs were scattered among the several boroughs mostly located in outer London. Many LSOAs with lower life expectancies were clustered in the eastern boroughs of Greenwhich, Newham, Tower Hamlets, and Lewisham. For males this diagonal cluster was more pronounced and extended further. (b) Estimated increase in life expectancy between 2002 and 2024. For females the largest increases were in the western boroughs of Richmond on Thames and Hounslow and the northern boroughs of Haringey and Waltham forest. For males, however, the increases were more concentrated in the central-eastern boroughs of Tower Hamlets and Islington.}
    		\label{fig:Fittede0}
    	\end{figure}

    	At a fine spatial resolution, Figure~\ref{fig:Fittede0} presents our estimates for all LSOAs by sex. The top panels (a) shows LSOA-specific life expectancy in 2024, while the bottom panels (b) illustrates the change between 2002 and 2024. These estimates clearly highlight improvements and provide much more detailed information than life expectancy computed directly from raw LSOA-level data (see Figure~\ref{fig:data}). In particular, the 2024 maps reveal clear spatial gradients, with patterns along both east-west and center-periphery axes. For both males and females, the areas with the highest life expectancy were clustered in central London, particularly in the boroughs of Kensington and Chelsea and Camden. This cluster of high life expectancy LSOAs extended across the Thames into Richmond upon Thames and further north into the borough of Barnet, where many of the highest estimated life expectancies were observed. In contrast, the lowest life expectancy LSOAs were concentrated along a diagonal axis in central eastern London, encompassing the boroughs of Lambeth, Southwark, Tower Hamlets, Newham, and Waltham Forest. Additional pockets of relatively low life expectancy were also found in the outer west (Hillingdon and Hounslow) and outer northeast (Redbridge and Havering). These spatial patterns were consistent for both males and females. The Spearman correlation between male and female life expectancy in 2024 was 0.74, confirming the overall similarity of geographic gradients between sexes.
    	
    	While the spatial distribution of life expectancy in 2024 is generally similar between sexes, differences become apparent when examining changes over time (bottom panels in Figure~\ref{fig:Fittede0}). The largest gains occurred in LSOAs in central eastern London for both sexes, but the areas of greatest improvement were shifted slightly eastward for males and northward for females. Notable increases were also observed in western LSOAs along the Thames. Overall, gains were more pronounced for males, with the largest estimated increase reaching 6.7 years in three LSOAs in Tower Hamlets. In contrast, LSOAs in the outer southeast and outer northwest experienced the smallest gains. Across all LSOAs, life expectancy increased relative to 2002 levels, with even the smallest gains exceeding approximately 2 years.
    	
\begin{figure}
    		\centering
    		\includegraphics[width=\textwidth]{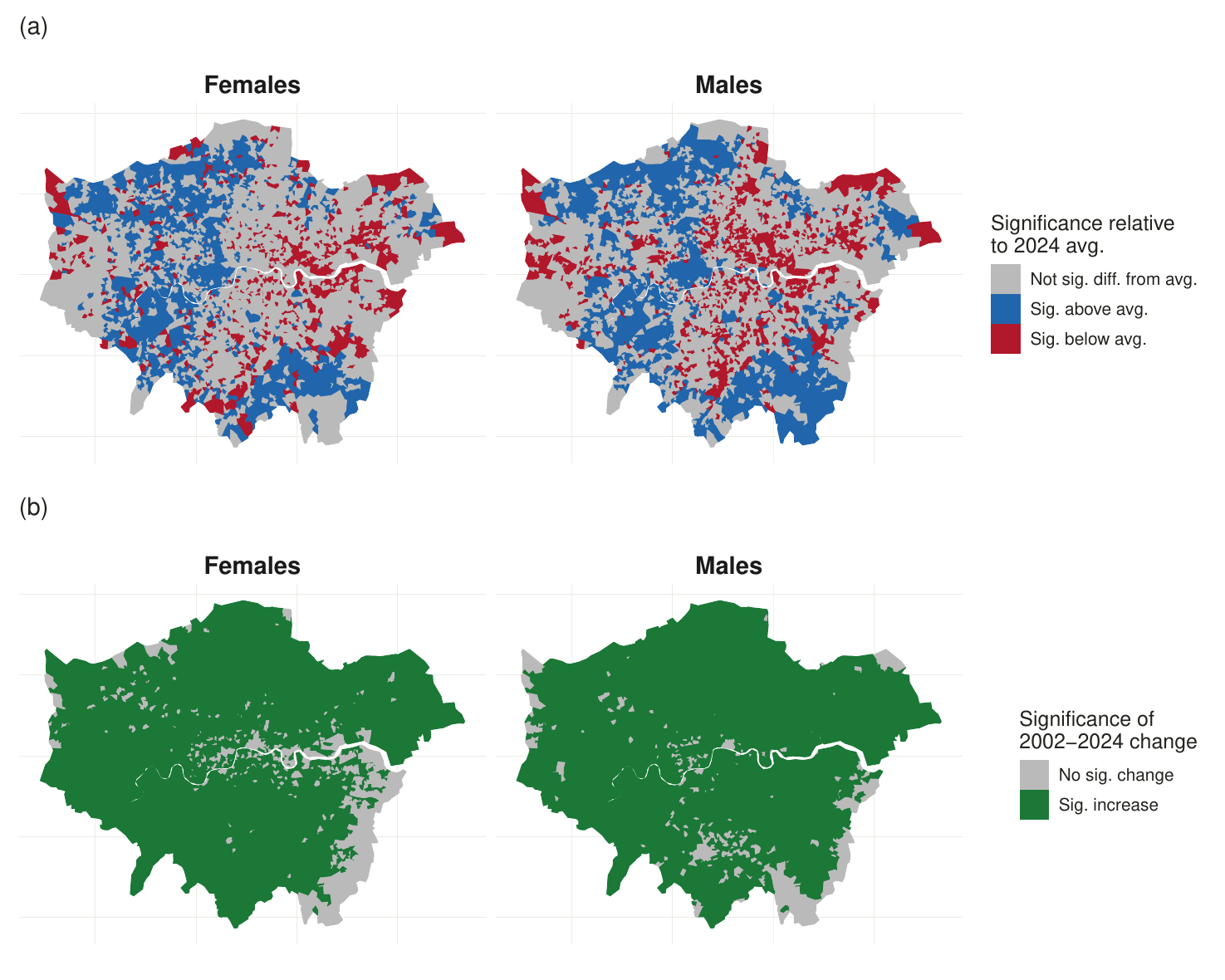}
    		\caption{(a) Spatial distribution of the statistical significance of the difference between our LSOA-level life expectancy estimates and the overall London average in 2024, stratified by gender. Gray indicates areas with no significant difference, while red and blue highlight areas where life expectancy is significantly below or above the 2024 average, respectively. The analysis confirms that much of the spatial variation observed, such as the broad western and central-eastern advantages and the disadvantages in outer western and eastern areas, is statistically significant. Notably, areas with higher life expectancy tend to be clustered, and the model captures meaningful differences within boroughs and among neighboring LSOAs. (b) Spatial distribution of the statistical significance of the change in life expectancy between 2002 and 2024 at the LSOA level, stratified by gender. Green denotes regions with a statistically significant increase, indicating areas where life expectancy has improved substantially over the period, while gray areas show no significant change. Most areas show statistically significant increases over time, reflecting overall improvements in health outcomes. However, outer southeastern LSOAs with the smallest estimated gains did not see significant increases, indicating persistent disparities.}
    		\label{fig:e0Significance}
    	\end{figure}

    	A key advantage of our approach is that it enables straightforward quantification of uncertainty both in the model estimates and in any derived quantities. To illustrate this, Figure~\ref{fig:e0Significance} visualizes the significance of our life expectancy estimates at the 95\% level, complementing the results shown in Figure~\ref{fig:Fittede0}. LSOA-specific confidence intervals were computed using a parametric bootstrap. To assess differences relative to the entire GLA, we also estimated confidence intervals for London-wide life expectancy using a 2D age-time smoothing model that accounts for the shock years 2020 and 2021 (i.e.~a version of our model without spatial components). The statistical significance of the life expectancy gradient in 2024 was evaluated by checking whether the 95\% confidence intervals for individual LSOAs overlapped with the 95\% confidence interval for the entire GLA. Similarly, to assess the significance of changes in life expectancy between 2002 and 2024, we examined, for each LSOA, whether the 95\% confidence intervals in 2024 overlapped with those in 2002.
    
    	We confirm that much of the spatial variation in life expectancy shown in Figure~\ref{fig:Fittede0} is statistically significant. The broad western advantage and central–eastern advantage are clearly evident, as is the disadvantage of outer western and eastern LSOAs. Figure~\ref{fig:e0Significance} further demonstrates that our model is sufficiently flexible to capture significant variation in life expectancy within the same borough and even among neighboring LSOAs. Regarding changes over time, increases in life expectancy are statistically significant for the vast majority of LSOAs. However, in outer southeastern LSOAs with the smallest estimated gains, life expectancy in 2024 was not significantly higher than in 2002.
    	
    	\subsection{The heterogeneous impact of Covid-19}
    	
\begin{figure}
    		\centering
    		\includegraphics[width=\textwidth]{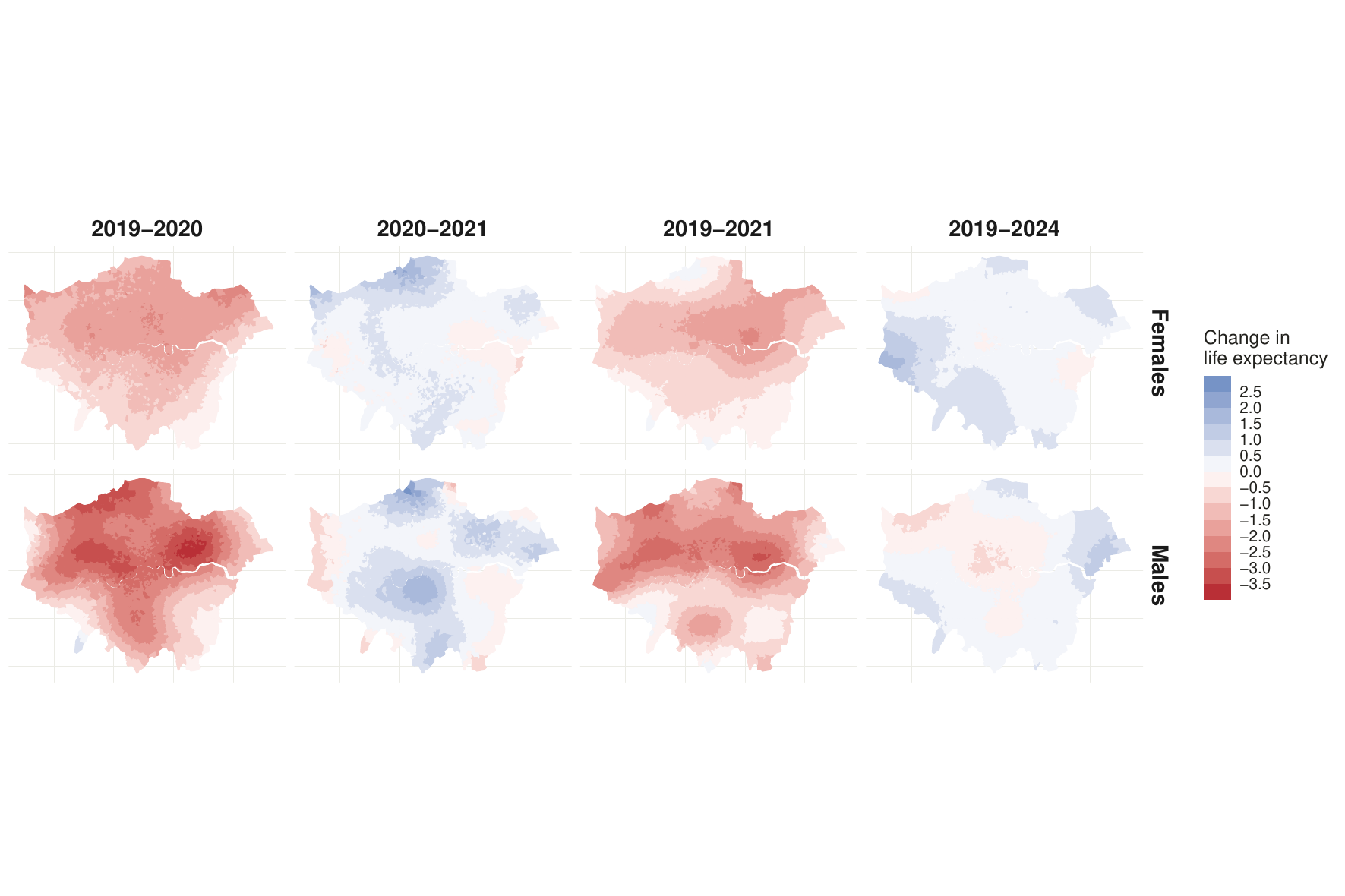}
    		\caption{Estimated change in LSOA-specific life expectancy between years during and after the Covid-19 mortality shock. In 2020, Covid-19 led to a drop in life expectancy in almost all areas, with the worst declines in the boroughs of Newham, Kensignton and Chelsea, and Brent. In 2021, life expectancy improved in most areas, but still remained below 2019 levels for the majority of LSOAs. By 2024, life expectancy was back at or above 2019 levels in most LSOAs, but notably had not fully recovered for males in many areas in central London.}
    		\label{fig:e0Change} 
    	\end{figure}

    	Focusing only on 2024 and long-term changes since 2002 risks overlooking short-term dynamics, most notably the profound disruption of the Covid-19 pandemic and the potential declines in life expectancy during the pandemic years. Of particular importance is the heterogeneous impact of Covid-19 across LSOAs in 2020 and 2021, as well as the extent to which mortality subsequently recovered to 2019 levels or realigned with the pre-pandemic trend. To explore these dynamics, Figure~\ref{fig:e0Change} presents changes in life expectancy over four distinct periods: 2019-2020, capturing the immediate impact of the pandemic; 2020-2021, highlighting shifts within the pandemic period; 2019-2021, reflecting the effect of two pandemic years; 2019-2024, assessing recovery relative to pre-pandemic mortality levels.	
    	
    	The impact of the 2020 shock on life expectancy was highly heterogeneous across London. Among males, the steepest declines were concentrated in LSOAs within Newham, where losses reached nearly 4 years, with another cluster of substantial decreases in Kensington and Chelsea and Brent. Strikingly, these western areas had some of the highest estimated life expectancies in 2019. For females, the largest drops—up to 2.5 years—were observed in Havering and Hackney. By 2021, overall life expectancy had risen relative to 2020, though it remained well below 2019 levels, and several LSOAs continued to experience declines. For males, further decreases were most evident in Hillingdon and Enfield, while for females they persisted in Havering and Hackney. When comparing 2024 with 2019, life expectancy in some central London LSOAs, particularly around Westminster, was still 0.4–0.7 years lower than before the pandemic. For females, the largest sustained deficits were more modest, ranging from 0.1 to 0.3 years, concentrated in Bexley in the east.
    	
    	\subsection{Age-specific mortality inequalities}
    	
\begin{figure}
    		\centering
    		\includegraphics[width=\textwidth]{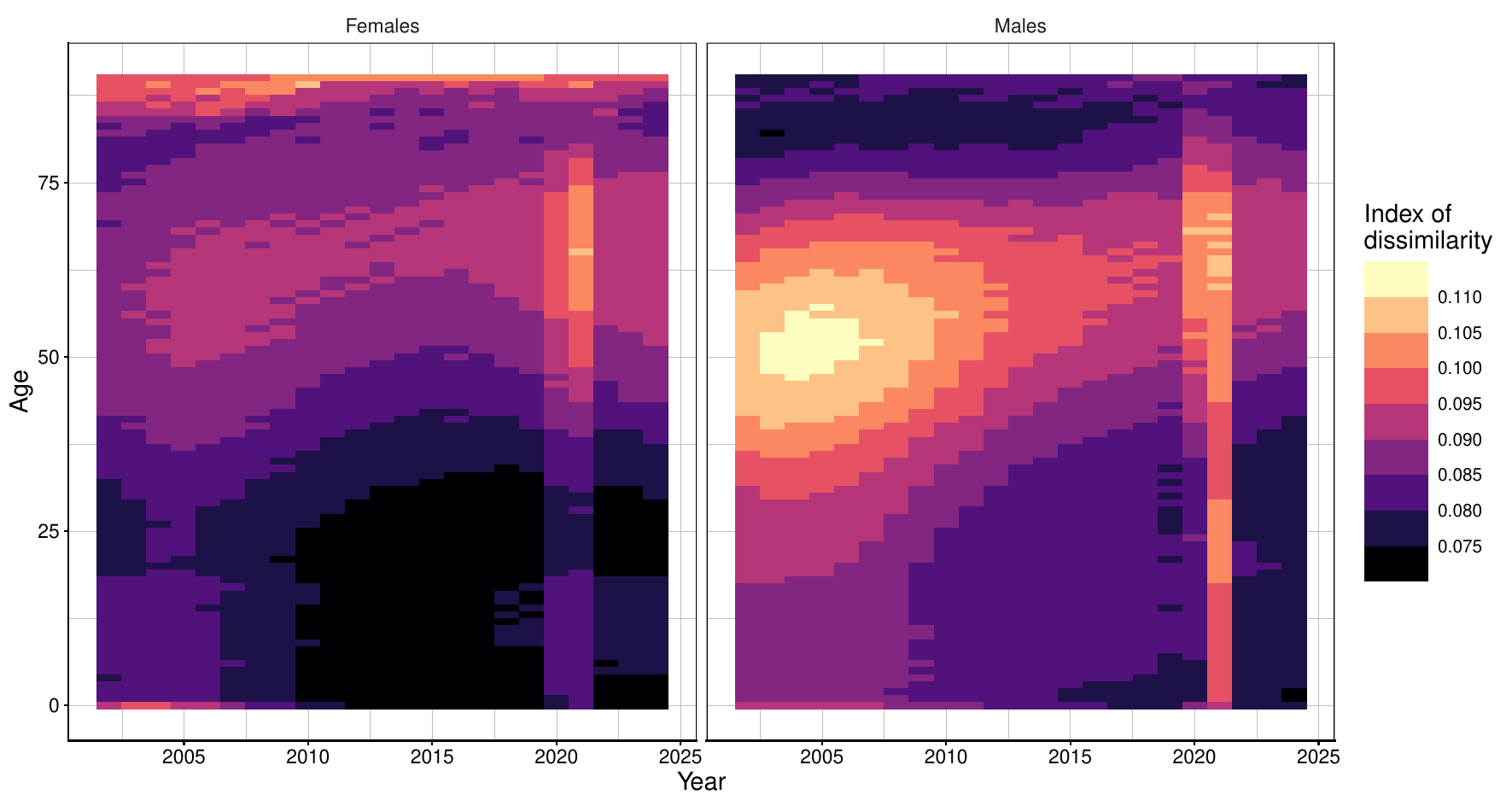}
    		\caption{Index of dissimilarity (ID) for mortality among London LSOAs by age, year, and sex. The index of dissimilarity represents the proportion of estimated deaths that would have to be redistributed to equalize the distribution of estimated deaths over LSOAs to the distribution of population exposed over LSOAs. For males, the index of dissimilarity shows a peak around age 50 in 2005, and for females, the highest values are for older ages and ages between 55 and 75. The Covid-19 shock years of 2020 and 2021 show clear increases in the ID. For females, a recent upward trend in the ID at younger ages appears to have begun before the pandemic and continued thereafter.}
    		\label{fig:ID}
    	\end{figure}

    	We now turn to results on trends in age-specific mortality inequalities. A key advantage of our approach is that for each LSOA we estimate a smooth age pattern of mortality, allowing us to compute indicators of inequality at every single age and year in the study period. For this purpose, we use the index of dissimilarity (ID), which measures the proportion of deaths that would need to be redistributed across LSOAs to equalize each area’s share of deaths with its share of the population, for a given age and year. Details on its computation and properties are provided in \textcite{WagstaffEtAl1991, MartinEtAlEJP2025}. Figure~\ref{fig:ID} displays the estimated ID across age and time, with darker colors indicating lower levels of inequality. In general, dissimilarity was higher for males than for females, though at older ages females exhibited a relatively higher ID. Among males, the highest levels of inequality occurred at midlife (around age 50) at the beginning of the study period. The effect of Covid-19 is also evident: the ID rose sharply across all ages in 2020 and 2021, revealing how the pandemic dramatically widened spatial divides in mortality, even within a single, albeit large, city. Although the ID subsequently declined after 2021, for females it rose again at younger ages (below 55) in 2023 and 2024, pointing to the emergence of possible longer term changes in mortality inequalities in London.
    	
    	\subsection{Model validation}
    	
    	Validating models of small-area mortality risk is inherently challenging, as there is never any ``ground truth'' against which estimates can be compared. Mortality data at this scale are sparse because the risk of death is low and local populations are small, we thus cannot appeal to additional data sources to know the true underlying age-year-LSOA-specific mortality risk. One option is to conduct simulation studies by treating the observed data as realizations of the stochastic process defined by our model. However, this approach merely assesses how well the estimation procedure can recover the model used to generate the data, thereby implicitly assuming that the model itself is an adequate approximation of reality. Other studies \parencite{AlexanderZagheniBarbieri2017, Denecke2023} have simulated small-area mortality data using alternative or external models, and then evaluated how well their methods could recapture the imposed ``ground truth''. Yet this approach is also inadequate, as it substitutes one pre-specified ``ground truth'' for another, while the true underlying mortality process remains fundamentally unobservable.
    	
    	Nevertheless, it remains essential to validate our model, both to demonstrate its reliability and to establish confidence in its practical applicability, and for this purpose we draw on an idea from the classic small-area estimation literature \parencite{Ghosh2020}. Direct estimates from raw data for larger areas, such as the entire GLA or its 33 individual boroughs, are generally sufficiently stable. Ideally, our small-area estimates at the LSOA level, when aggregated, should closely approximate these direct estimates for the larger areas. To validate our model, we therefore assess how well the aggregated fitted LSOA-level estimates reproduce borough-level direct estimates. Importantly, borough membership is not explicitly incorporated in our model. While age-specific estimates at the borough level can be somewhat unstable due to small population sizes and many zero counts at younger ages, life expectancy trends are sufficiently stable over time to provide a meaningful benchmark for comparison with our aggregated modeled estimates.
    	
\begin{figure}
    		\centering
    		\includegraphics[width=\textwidth]{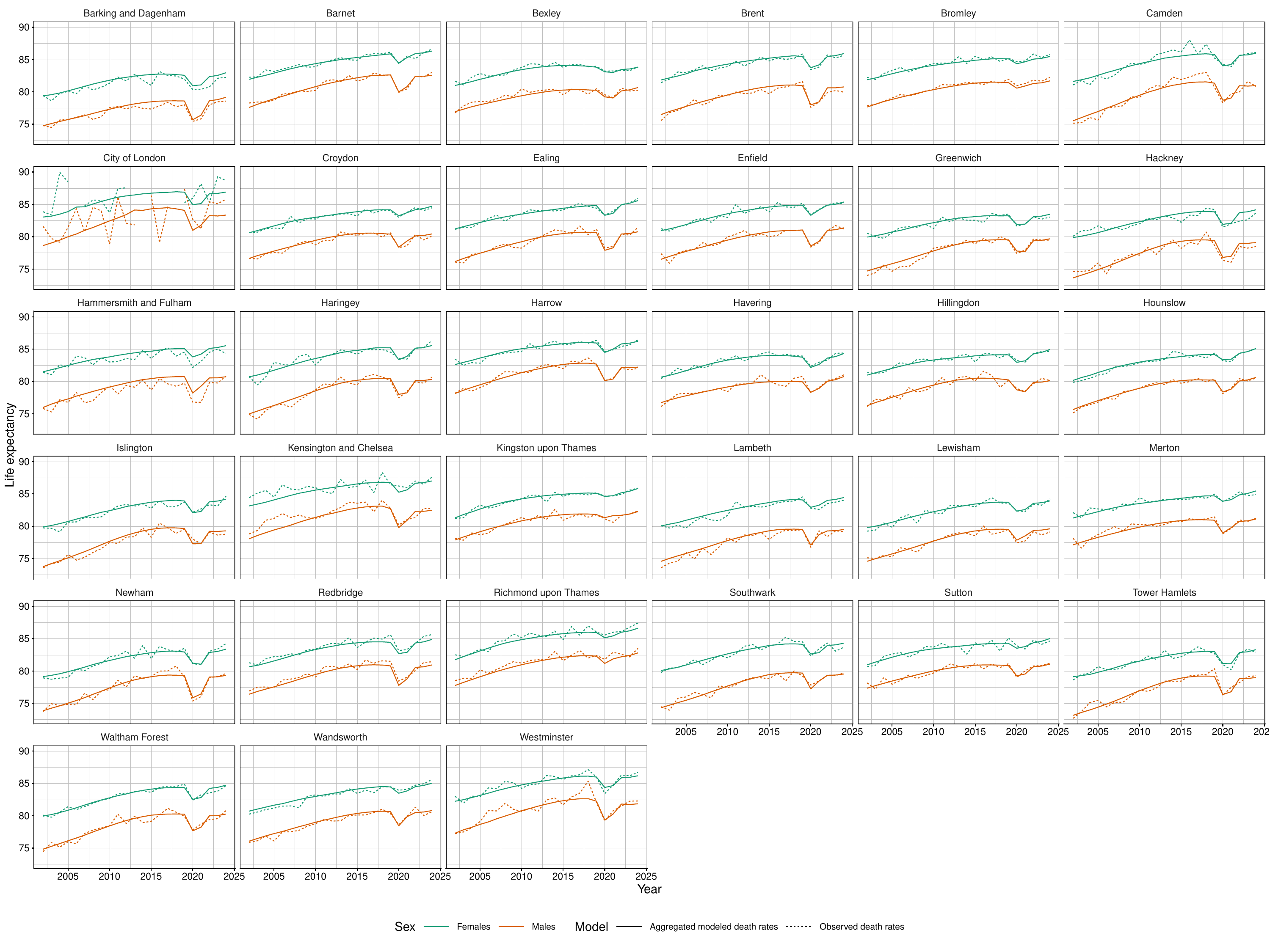}
    		\caption{Comparison of life expectancy estimates across the 33 London boroughs to evaluate model performance. The figure shows raw, direct estimates from observed data, which can be highly variable and include extreme values, and model-based estimates derived at the LSOA level and aggregated to the borough level. The model results in smoother estimates and, by construction, does not incorporate borough-level information. The y-axis is truncated at 95 years for clarity, which leads to the omission of the upper extreme for the very small population of the City of London, where estimates occasionally exceed this limit, such as over 100 years for men in 2018.}
    		\label{fig:AggregationDiff}
    	\end{figure}

    	Figure~\ref{fig:AggregationDiff} compares modeled and direct life expectancy values over time for each borough and sex. Direct estimates exhibit notable year-to-year fluctuations due to sampling variability, so an exact match is neither expected nor desirable, as this would indicate overfitting. Nonetheless, clear temporal trends are evident, and it is important that the aggregated modeled life expectancy capture these patterns, demonstrating that our LSOA-level results are consistent with observed data. Overall, the modeled estimates closely track borough-level trends, with no evidence of systematic bias. Some exceptions exist: in Camden, modeled life expectancy is slightly higher than observed until 2010 and lower until 2019, while in Hammersmith and Fulham, it remains consistently above direct estimates, indicating minor bias.
    	
    \section{Discussion}\label{sec:discussion}
        
        \subsection{Model strengths and weaknesses}
        
        Our modeling approach is based on the principle of simultaneously utilizing information from the aggregated data across all modeled areas while borrowing strength over age, space, and time. Conceptually, it extends the two-dimensional age-time $P$-spline smoothing technique commonly employed in mortality research, but with additional structure: it incorporates three additive components and allows for nonsmooth effects. Unlike models in the demographic literature \parencite{AlexanderZagheniBarbieri2017, SchmertmannGonzaga2018}, we do not impose an externally specified baseline age pattern of mortality. Instead, we estimate a standard smooth age-time mortality surface directly from the aggregate data, incorporating its associated uncertainty into the model. Moreover, our approach captures the full interaction among age, space, and time, providing flexibility through spline-based estimation. In our application to data including the Covid-19 pandemic years, we demonstrate the model's capacity to easily incorporate mortality shocks, highlighting its adaptability to real-world epidemiological events.
        
        While our approach offers considerable flexibility and can be applied across various contexts, it does have certain limitations. Our model leverages information from neighboring spatial units and the aggregate of all units to estimate unit-specific mortality schedules. However, when the combined population of these areas is too small to reliably estimate age-specific hazards, the model's performance may decline. In cases of severe data scarcity, it may be advantageous to consider methods that incorporate external structural information, such as model life tables (originating with \textcite{CoaleDemeny1966} and further developed in subsequent research) or TOPALS \parencite{DeBeerTOPALS2012}, which rely on external data. Alternatively, future extensions of our approach could be developed to similarly integrate such external sources, thereby enhancing reliability. Additionally, our model incorporates spatial information using the centroids of the units and does not account for alternative spatial structures, such as length, borders, or other relational configurations, which could be considered in future research. We chose this approach to maintain a parsimonious framework based on the 2D age-time $P$-spline smoothing. Incorporating hierarchical or alternative spatial structures could further enhance model accuracy, particularly in contexts where such spatial relationships are of primary interest beyond mere proximity.
        
        \subsection{Comparison with existing work on London}
        
        To contextualize our findings and assess their consistency with existing research, we compare our estimates with those from a previously published LSOA-level study by \textcite{BennettEtAl2023}. In the supplementary material, we present maps of life expectancy in 2019 and the changes from 2002 to 2019. While our results show a similar spatial gradient for life expectancy in 2019, the range of our estimates is narrower. The key difference lies in the patterns of change over time. \textcite{BennettEtAl2023} reported the largest increases in central-western London-areas with the highest life expectancy in 2019, whereas we found the greatest improvements further east and north, outside the central zones.  Our estimated increases are also smaller; the maximum increase we identified is 6.7 years for males, compared to over 10 years in some areas reported by \textcite{BennettEtAl2023}. Additionally, we found no areas with a decline in life expectancy, in contrast to their findings of some decreases. This highlights notable differences between our estimates and those of \textcite{BennettEtAl2023} in both the magnitude and spatial distribution of life expectancy changes.  
        
        \subsection{The impact of Covid-19}
        
        An additional contribution of our study is the ability to analyze mortality dynamics during the Covid-19 pandemic years. Previous research has uncovered substantially higher Covid-19 mortality risk in London compared to the rest of England and Wales \parencite{Feng2023}, particularly during the first wave \parencite{Adin2022}. Our high-resolution analysis reveals that, in 2020, a cluster of elevated mortality emerged in east London, an area where life expectancy was often significantly below the 2019 average. Other notable high-mortality clusters in 2020 were located in the west, where several LSOAs had significantly above-average life expectancy in 2019, and south of the Thames, where life expectancy tended to be closer to London's overall average. These findings indicate that during the initial wave of the pandemic, the areas experiencing the highest mortality shocks did so independently of their pre-pandemic mortality advantage or disadvantage. This pattern aligns with existing research on spatial disparities in Covid-19 mortality across Europe, which found that excess mortality in 2020 was not strongly correlated with pre-pandemic life expectancy levels \parencite{Bonnet2024}. In 2021, mortality rates improved relative to 2020, particularly south of the Thames; however, the three primary mortality clusters persisted. Furthermore, although our estimates suggest that life expectancy in London recovered to 2019 levels by 2024, some LSOAs failed to rebound—most notably in the central-western area around Westminster. Remarkably, these are areas that had some of the highest life expectancy levels in 2019. This counterintuitive result, where the most advantaged areas did not fully recover, raises important questions about differential resilience. On the other hand, preliminary research at the national level indicates that, in some southern and northern European countries with very high pre-pandemic longevity, life expectancy deficits persisted through 2024 \parencite{Dowd2025}. 
        
    \section{Conclusion}\label{sec:conclusion}
        
        In this paper we introduced a model of mortality in small areas that borrows strength over age, space, and time, as well as the total aggregate population being modeled. The model is based on a penalized spline methodology and is a natural extension of widely used smoothing techniques for mortality data. The modeling framework is highly flexible and can incorporate mortality shocks. In our application to over 4800 extremely small areas in the Greater London Authority between 2002 and 2024, we uncovered a unequal pace of increase in life expectancy, and a highly spatially heterogeneous impact of the Covid-19 shock on mortality and life expectancy. Furthermore, our estimates revealed large increases in age-specific mortality inequality due to Covid-19, as well as potentially longer-term increases emerging at younger ages since 2015. This versatile approach opens avenues for future research into spatial mortality inequalities across other urban settings or broader regions.
	
    \printbibliography

    \appendix
    
\end{document}